\documentclass[preprintnumbers,aps,prd,twocolumn,showpacs,showkeys,floatfix,preprintnumbers,letterpaper,amsmath,amssymb,superscriptaddress]{revtex4-1}
\usepackage{subfigure}
\usepackage{graphicx}
\usepackage{dcolumn}
\usepackage{epsfig}
\usepackage{amsmath, bm}
\usepackage{amsfonts}
\usepackage{amssymb}
\usepackage{color} 
\usepackage{xcolor}
\usepackage{hyperref}
\usepackage{tabularx}
\usepackage{float}
\usepackage[normalem]{ulem}

\usepackage[T1]{fontenc}
\usepackage{ae,aecompl}
\newcommand{\be}{\begin{equation}}
\newcommand{\ee}{\end{equation}}
\newcommand{\bea}{\begin{eqnarray}}
\newcommand{\eea}{\end{eqnarray}}

\begin{document}

\title{ \textbf{Optimal Transport Reconstruction of Biased Tracers in Primordial Non-Gaussian Fields} } 

\author{Farnik Nikakhtar}
\email{farnik.nikakhtar@yale.edu}
\affiliation{Department of Physics, Yale University, New Haven, CT 06511, USA}

\author{Ravi K.~Sheth}
\affiliation{Center for Particle Cosmology, University of Pennsylvania, Philadelphia, PA 19104, USA}

\author{Nikhil Padmanabhan}
\affiliation{Department of Physics, Yale University, New Haven, CT 06511, USA}
\affiliation{Department of Astronomy, Yale University, New Haven, CT 06511, USA}

\author{Bruno L\'evy}
\affiliation{Centre Inria de Saclay, Université Paris Saclay, Laboratoire de Mathématiques d'Orsay, Paris, France}

\author{Roya Mohayaee}
\affiliation{Sorbonne Universit\'e, CNRS, Institut d'Astrophysique de Paris, 98bis Bld Arago, 75014 Paris, France}
\affiliation{Rudolf Peierls Centre for Theoretical Physics, University of Oxford, Parks Road, Oxford OX1 3PU, United Kingdom}

\date{\today}

\begin{abstract}
Optimal transport provides an efficient method to infer the displacement of objects by mapping their initial positions to their present-day locations over cosmic time; equivalently, it enables the reconstruction of initial positions from measurements taken at later times.  The method has been shown to be accurate even if positions for only a biased subset of the particles are measured, provided that the initial displacement field was Gaussian. The method does not rely on the assumption of a Gaussian displacement field, and thus may be extended to the reconstruction of non-Gaussian initial conditions. Here, we demonstrate how this is achieved for a class of “local” primordial non-Gaussian fields of current interest in cosmology.  For these models, there is a distinctive signature in the large scale clustering of biased tracers which depends on the product of the primordial amplitude $f_{\rm NL}$ and the nature of the tracers $b_\phi$.  Our method exploits the fact that this signature is not present in the full field; it is only present in biased fields.   Therefore, the mass that is not in the biased subset, what we call the `dust', also has a characteristic scale-dependence, albeit of a different amplitude.  We show that the quality of the optimal transport reconstruction improves as the model for this dust becomes more realistic.  
\end{abstract}



\maketitle


\section{Introduction}\label{intro}

Optimal Transport (OT) provides a powerful framework for reconstructing the cosmic displacement field, with applications ranging from the distance scale to primordial velocity and density field  \cite{levysebroy2020,PRLhalos,OTrsd,OTmt}. In our previous works we have demonstrated that OT accurately recovers displacements of biased tracers under Gaussian initial conditions. A critical open question is whether this framework extends to non-Gaussian initial conditions - a timely challenge given mounting efforts to constrain primordial non-Gaussianity with next-generation surveys \cite{sphereX,fNLdesi,fNLplanckDESI}. Here we show that it does.

We study if OT reconstruction remains useful for non-Gaussian initial conditions as well.  We are optimistic because the OT method is based on minimizing a suitably defined `cost function' \cite{nature,eur03,mohayaee2006}, but this cost does not assume Gaussianity; moreover, on the large scales that are most relevant to primordial non-Gaussianity, reconstruction is expected to be relatively mild.  On the other hand, our semi-discrete implementation of OT \cite{journals/M2AN/LevyNAL15} makes heavy use of the assumption that the initial spatial distribution was uniform \cite{PRLdm,PRLhalos}. Although this is almost certainly a good assumption for the full dark-matter field, it is not true for biased subsets of the full field \cite{st1999,PRLhalos}.  In practice, this means that we must use the spatial distribution of observed tracers to build a model of the subset of the field that was not observed, which we refer to as the `dust'.  How one models the dust may need to depend on assumptions about the initial conditions. 

In this paper, we focus on a particular model for primordial non-Gaussianity, in which the potential is a power series in a single Gaussian field:  
\begin{equation}
    \phi_{\rm NG}\equiv \sum_j f_{{\rm NL}_j} 
    \frac{\phi^j_{\rm G} - \langle\phi_{\rm G}^j\rangle}{j!} \approx 
     \phi_{\rm G} + f_{\rm NL} (\phi_{\rm G}^2 - \langle\phi_{\rm G}^2\rangle),
     \label{eq:fnl}
\end{equation}
which we truncate at $j=2$.  This is often refer\cite{}red to as the local-PNG model.  
Clearly, $f_{\rm NL}\ne 0$ controls the departures from Gaussianity.  In this model, 
\begin{align}
  \delta_{\rm NG} &\equiv -\nabla^2\phi_{\rm NG}
  = -\nabla^2 [\phi_{\rm G} + f_{\rm NL} (\phi_{\rm G}^2 - \langle\phi_{\rm G}^2\rangle)]\nonumber\\
  &= -\nabla^2 \phi_{\rm G} - 2f_{\rm NL} [\phi_{\rm G}\nabla^2 \phi_{\rm G} + |\nabla\phi_{\rm G}|^2]\nonumber\\
  & = \delta_{\rm G}\, [1 + 2f_{\rm NL}\,\phi_{\rm G}] -  2f_{\rm NL}\,|\nabla\phi_{\rm G}|^2
 \label{eq:deltaNL}
\end{align}
Since $\delta_{\rm NG}$ depends on the product of Gaussian fields $\delta_{\rm G}$ and $\phi_{\rm G}$, it is non-Gaussian.  One could study this using the higher order cumulants.  However, previous work \cite{PNGdalal} has shown that in such models, the statistics of the power spectrum of biased tracers reveal a qualitatively new and distinct feature at small $k$, which a number of surveys aim to exploit.  

In Section~\ref{sec:bk}, we study this particular case and highlight the fact that the dust, too, picks up this feature.  Section~\ref{sec:dust-k} describes how -- with one additional free parameter -- we can extend the Gaussian dust model to incorporate $f_{\rm NL}\ne 0$.  It also shows that, with this one change, the OT methodology goes through smoothly and accurately.  A final section summarizes.

We illustrate our methods and results using measurements in the Quijote-PNG \cite{QuijotePNG} simulation set.  All our results are averaged over 20 simulation boxes.  Each box is periodic with side $L=1h^{-1}$Gpc and contains $512^3$ equal mass particles.  The background cosmological model is flat $\Lambda$CDM with $(\Omega_m,\Omega_b,h) = (0.3175, 0.049, 0.6711)$.  For this model, the linear theory growth factor at $z=0$, normalized so that it scales as $D(z) = 1/(1+z)$ at early times, is $D(0) = 0.778$.  
The initial conditions were laid down at $z=127$, with a power spectrum $P_{\rm Lin}(k)$ whose shape and amplitude parameters are $(n_s,\sigma_8)=(0.9624,0.833)$.  
The biased tracers we consider are halos more massive than $20\,  m_p$ (where $m_p = 6.5\times 10^{11}h^{-1}M_\odot$), identified in the $z=0$ outputs with a FOF finder (see Ref.\cite{OTrsd} for further details about how the halos were identified).  In all the clustering measurements which follow, we weight each object by its mass.  The mass fraction that is bound up in these biased tracers is $f_b=0.22$.

\section{Scale-dependent bias}\label{sec:bk}

\begin{figure}
    \centering
    \includegraphics[width=\linewidth]{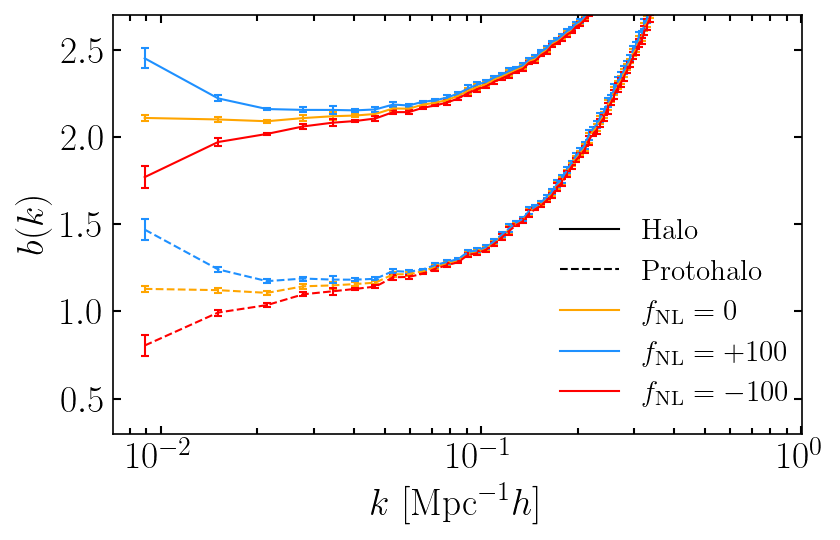}
    \caption{Scale dependent bias of halos and their corresponding protohalos, for three choices of $f_{\rm NL}$.  At small $k$, the two sets of curves differ by an additive constant: unity.  
    \label{fig:bk}}
\end{figure}

It is known that, in such models, 
\begin{equation}
    b(k) \equiv \frac{\delta_b(k)}{\delta_m(k)} 
    \approx b_\delta + 2\, \frac{f_{\rm NL}\,b_\phi}{\alpha(k,z)} ,
    \label{eq:bk}
\end{equation}
where $\delta_b$ and $\delta_m$ are the tracer and matter fields, 
$\alpha(k,z) = D(z)\,(kc/H_0)^2/(3\Omega_m/2)$, 
and $D(z)$ is the linear theory growth factor normalized to $D(z) = 1/(1+z)$ at early times, when matter dominates \citep{fNLoptimal}.  Thus, $b(k)$ has a characteristic $k^{-2}$ scaling at $k\ll c/H_0$.  Figure~\ref{fig:bk} shows that this scaling is seen both for the halos and their corresponding protohalos.  

\begin{figure}
    \centering
    \includegraphics[width=\linewidth]{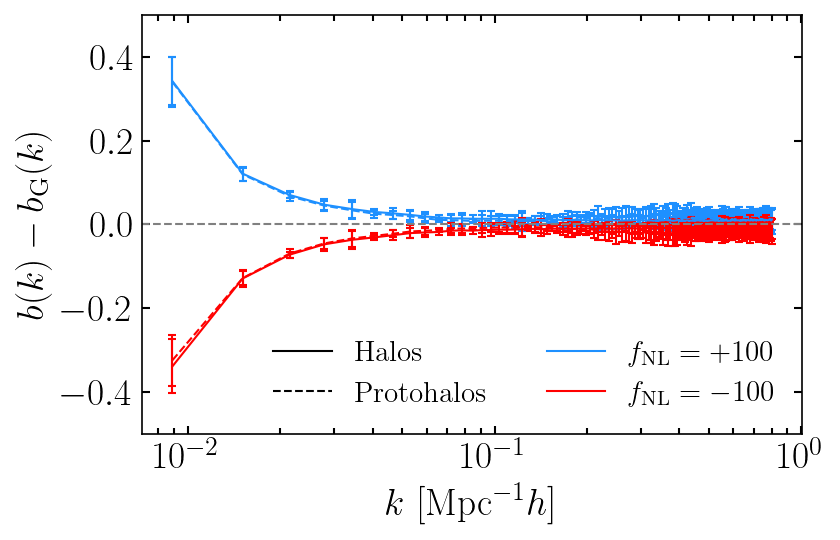}
    \caption{At small $k$, the difference from the $f_{\rm NL}=0$ bias is the same for halos as for protohalos:  the offset in Figure~\ref{fig:bk} is due to the evolution of $b_\delta$.  
    \label{fig:bk-bG}}
\end{figure}

Notice that the halo and protohalo samples appear to be offset by unity.  For Gaussian initial conditions, it is well known that $b_\delta^{\rm proto} = b_\delta - 1$ \cite[this is the Lagrangian-to-Eulerian bias relation of Refs.][]{mw1996,st1999,bkPeaks}.  Since evolution cannot affect large scales, we do not expect any qualitatively new effect when $f_{\rm NL}\ne 0$:  in particular, we do not expect the $k^{-2}$ term to evolve (Appendix~\ref{sec:conserved} provides more details).  To check this, Fig.~\ref{fig:bk-bG} shows $b(k) - b_{\rm G}\approx b(k) - b_\delta$.  This effectively removes the offset, so the similarity of the two sets of curves implies that $b_\phi^{\rm proto}\approx b_\phi$ (c.f. equation~\ref{eq:bk}).  

We only ever observe a biased subset of the full field.  We refer to the part of the field that is not observed as the `dust'.  In OT analyses, the observed tracers are always mass-weighted.  If the mass fraction that is observed is $f_b$, then the mass fraction in dust is $1-f_b$.  

\begin{figure}
    \centering
    \includegraphics[width=\linewidth]{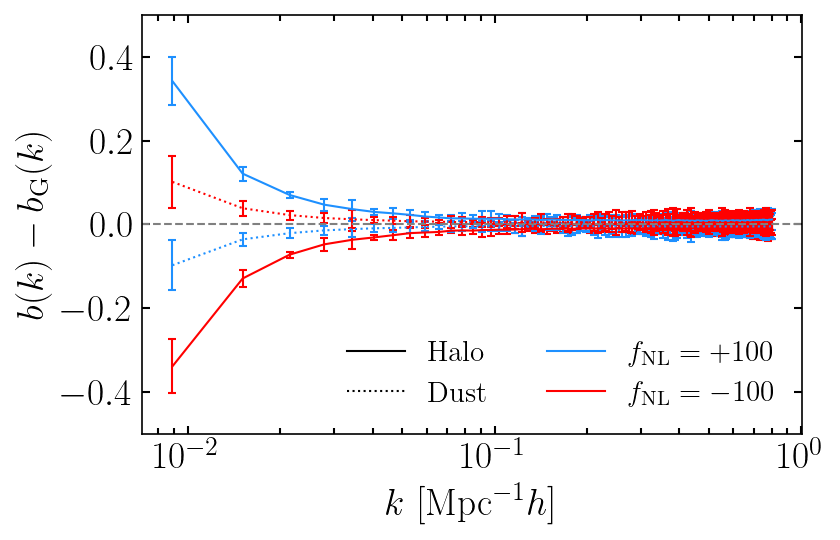}
    \caption{The bias of the `dust' -- the particles that are not in the halo catalog -- also exhibits $k$-dependent bias.  Its amplitude -- and the fact that the $k$-dependence is opposite to that of the halos -- is set by the requirement that the total field show no $k$-dependence (Eq.~\ref{eq:fb}).  
    \label{fig:bk-dust}}
\end{figure}

The first thing to note is that, to a good approximation, the full field does {\em not} exhibit $k$-dependent bias.  As the full field is just the sum of the mass-weighted halo and dust fields,  
\begin{equation}
    f_b\,b(k) + (1-f_b)\,b_{\rm dust}(k) = 1,
    \label{eq:btotal}
\end{equation}
where the right hand side is just the statement that the full field is not biased with respect to itself.  Therefore, 
\begin{align}
    f_b\,b_\delta + (1-f_b)\,b_{{\rm dust}\delta} &= 1
    \qquad {\rm and}\\
    f_b\,b_\phi + (1-f_b)\,b_{{\rm dust}\phi} &= 0.
    \label{eq:fb}
\end{align}
Hence, $b_{{\rm dust}\phi}\ne 0$:  the dust has its own $k$-dependent bias.  Moreover, since $f_b$ cannot exceed unity, $b_\phi$ and $b_{{\rm dust}\phi}$ must have opposite signs.  Fig.~\ref{fig:bk-dust} shows that this is correct.  In addition, we have checked that the bias of the dust evolves similarly to the bias of the tracers, meaning that 
 $b_{{\rm dust}\delta}^{\rm proto} = b_{{\rm dust}\delta} - 1$
and 
 $b_{{\rm dust}\phi}^{\rm proto} = b_{{\rm dust}\phi}$.  

\section{Scale-dependent dust modeling}\label{sec:dust-k}
For Gaussian initial conditions ($f_{\rm NL}=0$), a Wiener filter model produces a good estimate of the distribution of the dust from observations of the biased tracers alone \cite{OTrsd}.  This model rescales the amplitude of each observed Fourier mode while keeping the phase information unchanged:   
\begin{equation}
  \delta_{\rm dust}(\bm{k}) 
  = \frac{P_{db}(k)}{P_{bb}(k)}\,\delta_b(\bm{k})
  \approx \frac{b(k)b_{\rm dust}(k)}{b^2(k) + [\bar{n}P_{\delta\delta}(k)]^{-1}}\,\delta_b(\bm{k})
  \label{eq:Wdust}
\end{equation}
(where $P_{\delta\delta}(k)$ is the power spectrum of the full field, and the $\bar{n}$ accounts for the mass-weighting of the discrete biased tracers).  
Evidently, to implement this, we must guess values for $b_\delta, f_\text{NL}b_\phi$ and $b_{{\rm dust}\delta}, f_\text{NL}b_{{\rm dust}\phi}$.  However, if we guess $b_{\delta}$, $f_\text{NL}b_\phi$ and $f_b$, then Eq.(\ref{eq:fb}) fixes the parameters of the dust.  For Gaussian initial conditions, we must guess $f_b$ and $b_{\delta}$, so here we must guess one more parameter:  $2f_{\rm NL}b_\phi$.  Note that we do not need to know $f_{\rm NL}$ and $b_\phi$ separately.  Since $2f_{\rm NL}b_\phi$ can be estimated from the observed tracer, the data itself puts a prior on its value \cite[for the DESI survey, see Eq.~6.4 in Ref.][]{fNLdesi}.

\begin{figure}
    \centering
    \includegraphics[width=\linewidth]{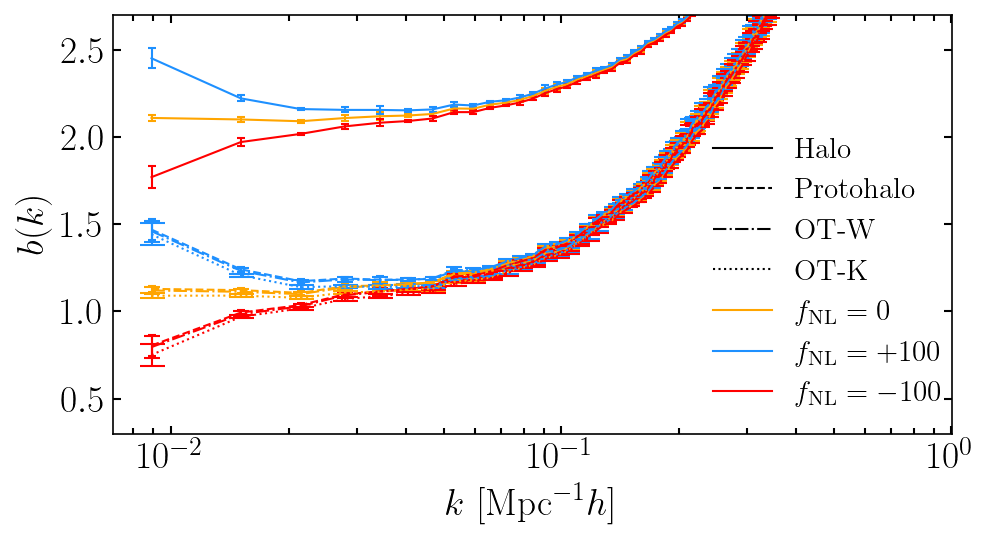}
    \caption{Scale dependent bias of the pre- (upper solid curves) and post-reconstructed halos: lower dotted curves are based on a Wiener filter model for the dust, and dot-dashed curves use the true dust.  Both are similar to the protohalos (dashed), though the true dust results are in slightly better agreement.
    \label{fig:bk-OTK}}
\end{figure}

In what follows, we refer to reconstructions based on the Wiener-filtered ($k$-dependent) dust model as OT-K.  The dotted curves in Fig.~\ref{fig:bk-OTK} show that OT-K correctly reconstructs the positions of the biased tracers in the sense that they exhibit the appropriate scale-dependent bias (short dashed curves show the scale dependence of the actual protohalos).  Some of the small differences between the OT-K reconstructions and actual protohalos arise because OT-K is based on an approximate (Wiener-filter) model for the dust.  Therefore, we have also reconstructed the field comprised of the observed biased tracers and (a random sample of) the true dust particles.  We refer to this as OT-W.  The dot-dashed curves in Fig.~\ref{fig:bk-OTK} show that OT-W is in even better agreement with the protohalos.  We will comment on the implications of this in the final discussion of this paper.

\begin{figure}
    \centering
    \includegraphics[width=\linewidth]{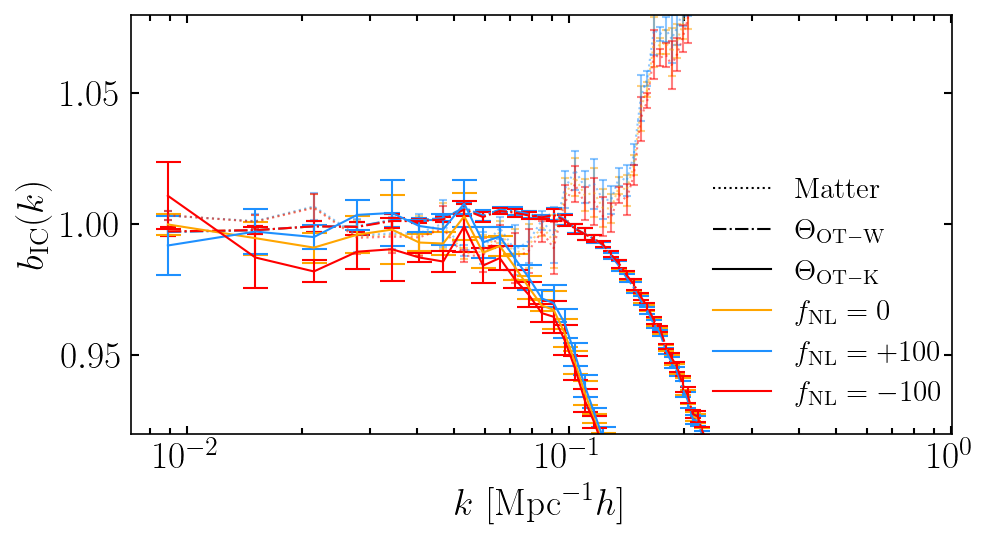}
    \caption{Bias of the full matter field pre- (dotted curves) and post-reconstruction (dot-dashed and solid); the latter use the divergence of the OT-W and OT-K displacement fields, evaluated at the reconstructed positions, as a proxy for the linear theory overdensity.  The better dust model (dot dashed) provides a factor of 2 improvement in terms of the $k$ values over which $b(k)\approx 1$.  
    This lack of any appreciable $k$ dependence here, along with the previous figure showing that the bias of the reconstructed tracers is similar to that of the protohalos, suggests that our method has worked well.
    \label{fig:bk-divS}}
\end{figure}

Rather than showing that the dust is also reconstructed accurately (it is), we instead consider the full reconstructed field, for which the divergence of displacements serves as a useful proxy \cite{OTmt}, at least when $f_\text{NL}=0$. That is, we evaluate  
\begin{equation}
  \theta_\text{OT}({\bm q}) \equiv \nabla_{\bm q} \cdot {\bm S}_{\text{OT}} ,
  \label{eq:thetaOT}
\end{equation}
where ${\bm S}_\text{OT}$ is the displacement vector between the initial position $\bm{x}$ and the reconstructed position $\bm{q}$ on a $512^3$ grid with a TSC mass assignment.  Fig.~\ref{fig:bk-divS} shows that the bias of this $\theta_\text{OT}$ field has almost no $k$-dependence (at small $k$), although this extends to larger $k$ for reconstructions based on OT-W (dot-dashed) than OT-K (solid).  As another measure of the (in)sensitivity to $f_\text{NL}$, the mean square value of 
$\langle\theta_\text{OT}^2\rangle = \int (dk k^2/2\pi^2)\, P_{\theta\theta}(k)$ equals $(1.68^2, 1.66^2,1.65^2)$ for $f_\text{NL} = (+100,0,-100)$, with a standard deviation of $0.06$ in all cases, when we sum over all the modes. Had we smoothed the field to reduce the impact of modes with $k\ge 0.1h$/Mpc, the differences would be smaller. 
For now, we simply note that Figures~\ref{fig:bk-OTK} and~\ref{fig:bk-divS} indicate that we have achieved our primary goal: our results show that the OT methodology is able to reconstruct the clustering, and hence should be useful for BAO analyses, even when the initial conditions are non-Gaussian.

\section{Discussion and conclusions}\label{sec:conclude}
We studied halos which form in simulations of primordial non-Gaussianity models of the form given by Equation~(\ref{eq:fnl}).  On large scales (small $k$), their clustering has a characteristic scale dependence (equation~\ref{eq:bk}).  Although the protohalo patches from which they form are clustered differently, their scale dependence is the same (Figures~\ref{fig:bk} and~\ref{fig:bk-bG}, and Appendix~\ref{sec:conserved}).  Thus, halos and protohalos are two tracer populations with the same scale dependence, even though their scale-independent bias is different.  This is a simple illustration of why one cannot assume a universal relation between the parameters $b_\delta$ and $b_\phi$ which appear in equation~(\ref{eq:bk}).

The clustering of the `dust' -- the particles that are not in halos -- also shows a characteristic scale dependence, whose amplitude is related to that of the halos (equation~\ref{eq:fb} and Figure~\ref{fig:bk-dust}).  We used this relation to inform a Wiener-filtered model for the dust (equation~\ref{eq:Wdust}).  We then used our Optimal Transport algorithm to determine the displacements which take the dust plus mass-weighted halos back to a uniform distribution.  The OT-reconstructed halo positions show very similar scale dependence to the actual protohalos (Figure~\ref{fig:bk-OTK}).  

In the OT-reconstructed field, the divergence of the OT displacements, evaluated at the reconstructed positions (equation~\ref{eq:thetaOT}), is a proxy for the initial density fluctuation field, linearly evolved to the present day.  On large scales (small $k$), the actual full initial field shows no scale-dependence; our OT-reconstructed proxy for the corresponding linearly evolved field also shows almost no scale-dependence (Figure~\ref{fig:bk-divS}), with better models for the dust yielding $b(k)\sim 1$ over a greater range of $k$ (compare dot-dashed with solid).  We conclude that, provided the model for the dust is sufficiently realistic, OT provides a useful framework for reconstruction even when the initial conditions are non-Gaussian.  We expect this conclusion to be generic, and not confined to the particular model we studied here (because Eq.~\ref{eq:btotal}, from which Eq.~\ref{eq:fb} followed, is generic).  Moreover, the fact that the OT-W based results are better than those based on OT-K (Figures~\ref{fig:bk-OTK} and~\ref{fig:bk-divS}) implies that there are gains to be had from improved modeling of the dust.  For example, iterative methods or machine learning-based approaches like the one in \cite{AEdust} may be useful here.  

In practice, the Wiener-filter dust model requires a guess for the value of the combination $f_\textrm{NL}b_\phi$.  Although this can be estimated directly from the data, the (survey specific) uncertainties in this estimate will propagate into the OT analysis which follows.  Quantifying their impact is the subject of work in progress.  

Although the local PNG signal targeted by surveys such as SPHEREx primarily resides on ultra-large scales that are already close to the linear regime, future constraints at the level $f_{\rm NL}\sim \mathcal{O}(1)$ will require increasingly precise control over tracer bias, nonlinear evolution, and hidden-matter systematics. Here, we have not aimed at demonstrating a direct improvement in $f_{\rm NL}$ constraints, but we have provided a framework within which reconstruction-based analyses can be consistently applied in PNG cosmologies. In particular, OT reconstruction may help connect observed biased tracers more closely to their primordial Lagrangian counterparts and reduce subtle nonlinear or stochastic contributions that become relevant in next-generation surveys. We therefore view this work as a foundational step toward incorporating reconstruction techniques into future high-precision PNG analyses.

\acknowledgements
FN acknowledges support from the Yale Center for Astronomy and Astrophysics Prize Postdoctoral Fellowship.
NP was supported in part by DoE DE-SC0017660. 
RKS is grateful to the members of the Astronomy Department at Yale for their hospitality in May 2024, and to the EAIFR and ICTP in summer 2024, when most of this work was completed. 
Finally, we acknowledge the hospitality of the Institute for Fundamental Physics of the Universe (IFPU) in Trieste, Italy, where part of this work was developed and presented during a group meeting/focus program in April 2025.

\appendix

\section{Conserved tracers}\label{sec:conserved}
Let $\delta_b(z)$ denote the overdensity of a set of tracers at redshift $z$, and let $\delta_m(z)$ denote the overdensity of the matter field at that same time.  Suppose that $\delta_b(z) = b(z)\,\delta_m(z)$, where $b(z)$, the constant of proportionality that relates the two is the `linear bias factor'.  If the set of tracers moves, tracers are neither created nor destroyed, then the continuity equation implies that the linear bias factors at the two times are related by 
\begin{equation}
  b(z_1)-1 = [b(z_0)-1]\, D(z_0)/D(z_1),
\end{equation}
where $D(z)$ is the linear theory growth factor (e.g. Eq.68 in \cite{bkPeaks}).
Suppose that $z_0=0$, and define 
\begin{equation}
    b_{\rm Lag} \equiv b(z_0)-1.
\end{equation}
Rearranging to isolate $b(z_1)$, and then multiplying by $\delta(z_1) = \delta(z_0)\,D(z_1)/D(z_0)$ yields 
\begin{equation}
  b(z_1)\, \delta(z_1) 
   = \left[b_{\rm Lag} + \frac{D(z_1)}{D(z_0)} \right]\,\delta(z_0) .
\end{equation}
Inserting Eq.~(\ref{eq:bk}) for $b$ in $f_{\rm NL}$ models yields 
\begin{align}
  &b_\delta(z_1)\, \delta(z_1) + 2f_{\rm NL} \, b_\phi\,\frac{3\Omega_m/2}{(ck/H_0)^2} \,\frac{\delta(z_1)}{D(z_1)}\\
          &= \left[b_{\rm Lag} + \frac{D(z_1)}{D(z_0)} \right]\,\delta(z_0)
           + 2f_{\rm NL} \, b_\phi\,\frac{3\Omega_m/2}{(ck/H_0)^2}\frac{\delta(z_0)}{D(z_0)},\nonumber
\end{align}
which shows that the scale-dependent piece (the term proportional to $b_\phi$) does not depend on $z_1$.  Figs.~\ref{fig:bk} and~\ref{fig:bk-bG} show that halos in simulations do indeed exhibit this behaviour.

\bibliography{refs}

\end{document}